\documentclass[twocolumn,trackchanges]{aastex7}
\usepackage{wasysym}
\usepackage[flushleft]{threeparttable}

\begin{document}

\title{Extreme anisotropies in deep layers of an exploding star: overabundance of Cr in the northeastern jet of Cassiopeia A} 
\author[0000-0002-6045-136X]{Vincenzo Sapienza}
\affiliation{INAF-Osservatorio Astronomico di Palermo, Piazza del Parlamento 1, 90134, Palermo, Italy}
\affiliation{Dipartimento di Fisica e Chimica E. Segr\`e, Universit\`a degli Studi di Palermo, Piazza del Parlamento 1, 90134, Palermo, Italy}
\email[show]{vincenzo.sapienza@inaf.it}
\correspondingauthor{Vincenzo Sapienza}

\author[0000-0003-0876-8391]{Marco Miceli}
\email{marco.miceli@inaf.it}
\affiliation{Dipartimento di Fisica e Chimica E. Segr\`e, Universit\`a degli Studi di Palermo, Piazza del Parlamento 1, 90134, Palermo, Italy}
\affiliation{INAF-Osservatorio Astronomico di Palermo, Piazza del Parlamento 1, 90134, Palermo, Italy}

\author[0000-0002-0603-918X]{Masaomi Ono}
\email{masaomi@asiaa.sinica.edu.tw}
\affiliation{Institute of Astronomy and Astrophysics, Academia Sinica, Taipei 106319, Taiwan, R.O.C}
\affiliation{Astrophysical Big Bang Laboratory (ABBL), RIKEN Pioneering Research Institute (PRI), 2-1 Hirosawa, Wako, Saitama 351-0198, Japan}

\author[0000-0002-7025-284X]{Shigehiro Nagataki}
\email{shigehiro.nagataki@riken.jp}
\affiliation{Astrophysical Big Bang Laboratory (ABBL), RIKEN Pioneering Research Institute (PRI), 2-1 Hirosawa, Wako, Saitama 351-0198, Japan}
\affiliation{RIKEN Center for Interdisciplinary Theoretical \& Mathematical Sciences (iTHEMS), 2-1 Hirosawa, Wako, Saitama 351-0198, Japan}
\affiliation{Astrophysical Big Bang Group (ABBG), Okinawa Institute of Science and Technology Graduate University, 1919-1 Tancha, Onna-son, Kunigami-gun, Okinawa 904-0495, Japan}

\author[0000-0002-8967-7063]{Takashi Yoshida}
\email{yoshida@yukawa.kyoto-u.ac.jp}
\affiliation{Yukawa Institute for Theoretical Physics, Kyoto University, Kitashirakawa Oiwake-cho, Sakyo, Kyoto 606-8502, Japan}

\author[0000-0001-5792-0690]{Emanuele Greco}
\email{emanuele.greco@inaf.it}
\affiliation{INAF-Osservatorio Astronomico di Palermo, Piazza del Parlamento 1, 90134, Palermo, Italy}

\author[0000-0003-2836-540X]{Salvatore Orlando}
\email{salvatore.orlando@inaf.it}
\affiliation{INAF-Osservatorio Astronomico di Palermo, Piazza del Parlamento 1, 90134, Palermo, Italy}

\author[0000-0002-2321-5616]{Fabrizio Bocchino}
\email{fabrizio.bocchino@inaf.it}
\affiliation{INAF-Osservatorio Astronomico di Palermo, Piazza del Parlamento 1, 90134, Palermo, Italy}

\begin{abstract}
Core-collapse supernovae drive nucleosynthesis under extreme thermodynamic conditions, and complex mechanisms are at work prompting the transport of heavy elements from deep stellar interiors into outer layers.
We present spatially resolved X-ray spectroscopy of Cassiopeia A’s (Cas A) northeastern (NE) jet using the archival 1 Ms Chandra/ACIS observations, and focusing on three fingers of the jet.
We report the highest Cr/Fe mass ratio (Cr/Fe $\sim0.14$) ever observed in Cas A, localized in a compact region within the southernmost finger in the NE jet.
Comparisons with nucleosynthesis models indicate that the NE jet originated approximately at the boundary separating the complete Si burning layer from the incomplete Si-burning layer.
We also find that mixing from different layers is needed to explain the chemical composition of the three fingers in the NE jet.
We also detect significant differences in the physical and chemical properties among the three fingers analyzed of the NE jet. 
In particular, we find that, unlike the other two, the southernmost finger originated from a slightly more peripheral region of the explosion.
Moreover, while the northern and central fingers lie almost in the plane of the sky, the southernmost finger is moving in a different direction, showing a velocity along the line of sight of $\sim2100$ km s$^{-1}$ towards the observer, with a tilt angle of $\sim16$\textdegree.
These findings highlight the NE jet’s role in ejecting material from the deepest explosive burning layers, providing new insights into the asymmetries originating in the inner layers of core-collapse supernovae.
\end{abstract}



\section{Introduction}\label{sec1}
Core-collapse supernovae (CCSNe) are among the most influential astrophysical events for driving the chemical enrichment of galaxies with elements formed through stellar and explosive nucleosynthesis processes, shaping the structure and dynamics of the interstellar medium, and accelerating cosmic rays. 
Nevertheless several aspects of the CCSNe explosion mechanism remain poorly understood,  particularly the complex physical processes occurring during the final stages of the progenitor’s collapse. In this context, asymmetries in the explosion event are especially valuable, as they provide indirect but crucial insights into the otherwise unobservable phases of the explosion and into the structure and nature of the progenitor star.

Cassiopeia A (Cas A) is a young supernova remnant (SNR), approximately $350$ years old, \citep{2001AJ....122..297T}, at a distance of about $3.4$ kpc \citep{1995ApJ...440..706R} originated from a CCSN explosion \citep{2008Sci...320.1195K}, and presents a unique opportunity to study the physics of these events.
Indeed, Cas A shows a complex morphology and a highly uneven distributions of ejecta, features that preserve vital clues about the supernova (SN) explosion and the progenitor's internal structure.
These features are likely due to the progenitor star losing most of its hydrogen envelope and resulting in a highly asymmetric explosion (e.g. \citealt{1996ApJ...470..967F,2006ApJ...645..283F,2016ApJ...818...17F,2024ApJ...965L..27M}).
This high degree of asymmetry is also an effect of the circumstellar medium \citep{2022ApJ...929...57V, 2022A&A...666A...2O, 2025A&A...696A.188O} and it is particularly evident in the spatial distribution of heavy elements within the remnant, as revealed by high-resolution X-ray observations from Chandra and XMM-Newton \citep{2002A&A...381.1039W,2004ApJ...615L.117H,2012ApJ...746..130H}.
These features reflect the underlying explosion mechanism, which is thought to involve strong asymmetries and non-spherical dynamics. 
Theoretical studies have explained important observational properties of Cas A with neutrino-driven explosion mechanism (\citealt{2017ApJ...842...13W,2021A&A...645A..66O,2025A&A...696A.108O}), but others observed features, jet-like structure in particular, were not reproduced, and their formation still remains an open issue.
Moreover, jet-like structures have been hypothesized to eject material from the deep layers of the progenitor star, bypassing regions where traditional mixing processes occur \citep{2006ApJ...644..260L}.
Such jet-like structures have also been observed in other remnants \citep{2018A&A...615A.157G,2021A&A...649A..56S,2017ApJ...834..189P,2008AdSpR..41..390M}, further highlighting their relevance in the context of CCSN explosions.

Understanding the composition and dynamics of these jets is therefore key to uncovering the physics of CCSNe.
Recently, Fe-peak elements such as Ti, Cr, and Mn have gained increasing attention as powerful diagnostics of asymmetric nucleosynthesis and explosion dynamics in CCSNe. 
Their distribution and chemical composition offer a direct window into the physical conditions of the burning regions. 
Early evidence for the nucleosynthetic link between Cr and Fe came from the Cr-K$\alpha$ emission line survey of young SNRs by \citet{2009ApJ...692..894Y}, who found a positive correlation between Cr and Fe ionization states, including Cas A, suggesting a common origin in explosive silicon burning. 

Subsequent work by \citet{2020ApJ...893...49S} detected Mn-K$\alpha$ emission in Cas A and reported a remarkably low Mn/Cr mass ratio, indicative of a higher neutron excess than predicted by standard models. 
This pointed to either an energetic or highly asymmetric explosion, possibly involving a subsolar-metallicity progenitor in a binary system.
Building on these findings, \citet{2021Natur.592..537S} provided the first observational evidence for high-entropy plumes in the southeastern part of Cas A showing the ejection of Fe-peak elements like Cr and Ti. These plumes were interpreted as products of $\alpha$-rich freeze-out, consistent with nucleosynthesis occurring under high-entropy, neutrino-driven convection in the innermost layers of the progenitor. 

More recently, \citet{2023ApJ...954..112S} confirmed that the Mn/Fe ratio in the southeastern part of Cas A requires significant contributions from neutrino interactions, providing robust evidence for neutrino-driven processes in the explosion mechanism.
In parallel, \citet{2022PASJ...74..334I} detected shocked stable Ti at the tip of Cas A’s northeast jet, showing a composition consistent with incomplete silicon burning. 
Their results suggest that the jet may have formed not as a direct product of the main explosion engine, but rather through a secondary, sub-energetic, and lower-temperature mechanism. 
However, their analysis treated the NE jet as a single, integrated structure, potentially overlooking localized variations in composition.

To bridge the understanding of CCSNe dynamics and nucleosynthesis pathways, it is crucial to study the morphology and the chemical composition in regions of strong asymmetry, such as the jets of Cas A. 
By exploiting the archival 1 Ms \textit{Chandra/ACIS} observation, we conduct a spatially resolved spectral analysis on the NE jet focusing on the 3 different regions. 
We characterize the X-ray emitting plasma, chemically and dynamically, of these structures and we compare their abundance pattern with nucleosynthesis models to understand their origin and the differences among them.

The Letter is organized as follows: in Sect. \ref{sect:data} we present the datasets and the data reduction process, whereas in Sect. \ref{sect:results} we show the results obtained with the spectral analysis. Discussions and Conclusion are drawn in Sect. \ref{sect:con}.

\section{Observation and Data Reduction}\label{sect:data}
We analyzed the archival 1 Ms \textit{Chandra}/ACIS-S observations of Cas A.
It is the deepest observation of Cas A, performed in 2004 from February 8$^{th}$ to May 5$^{th}$ (ObsIDs: 4634, 4635, 4636, 4637, 4638, 4639, 5196, 5319, and 5320; P.I.: U. Hwang; \citealt{2004ApJ...615L.117H}) with pointing coordinates $\alpha_{J2000}=$ 23$^h$ 23$^m$ 26.7$^s$ and $\delta_{J2000}=$ +58\textdegree 49' 03.0''.
Other \textit{Chandra}/ACIS observations are present in the archive. However, due to their rapid expansion, the positions of the ejecta vary across different epochs, potentially introducing errors in region selection. 
For this reason, we chose not to include data from other epochs.

Data were analyzed with \textit{CIAO} (v4.15), using \textit{CALDB} (v4.10.7), and reprocessed with the \texttt{chandra\_repro} task. 
For the image analysis, we used the \texttt{merge\_obs} command to generate fluxed and merged images and a merged event file.
The pile-up map was created using a counts image created using the \texttt{dmcopy} \textit{CIAO} tool from the merged event file and processed through the \texttt{pileup\_map} \textit{CIAO} tool.
Spectra have been extracted from the individual event files using \texttt{specextract} \textit{CIAO} tool, which also generates the corresponding ancillary response file (arf) and redistribution matrix (rmf). 
The spectra extracted from a single observation have been merged using the \textit{CIAO} tool \texttt{combine\_spectra}.
A background spectrum was extracted from a circular region in the same chip of the ACIS-S outside of the remnant, as shown in Fig. \ref{fig:extr}, \textit{lower right panel}. 
This background region is less contaminated by the emission form the source. Nevertheless, we verified that by selecting a different background region (e.g., a circle in the upper corner of Fig. \ref{fig:extr}, (\textit{lower right panel}) our results are unaffected, the best-fit values varying by much less than 1 $\sigma$. Spectra were grouped to have at least 25 counts per bin and analyzed using the software \textit{XSPEC} (v12.14.1; \citealt{1996ASPC..101...17A}) with the solar abundances table from \citet{1989GeCoA..53..197A} and AtomDB version 3.0.9. 
Since the source spectrum is much brighter than the background, it was subtracted before fitting.

\section{Results}\label{sect:results}
\subsection{Cr-K$\alpha$ emission line in the NE Jet}
\begin{figure*}[t!]
    \centering
    \includegraphics[width=\columnwidth]{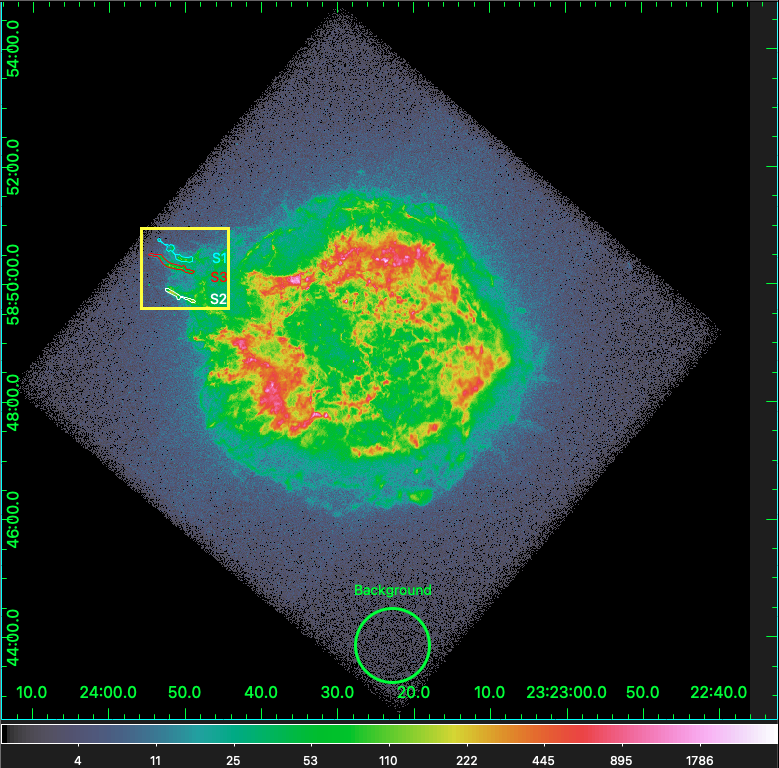}
    \includegraphics[width=\columnwidth]{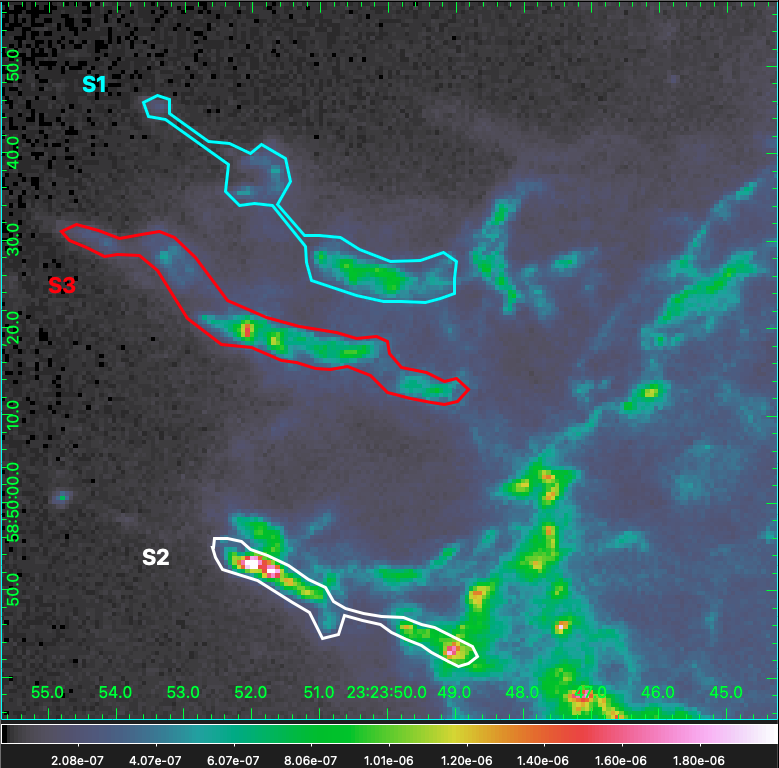}
    \includegraphics[width=\columnwidth]{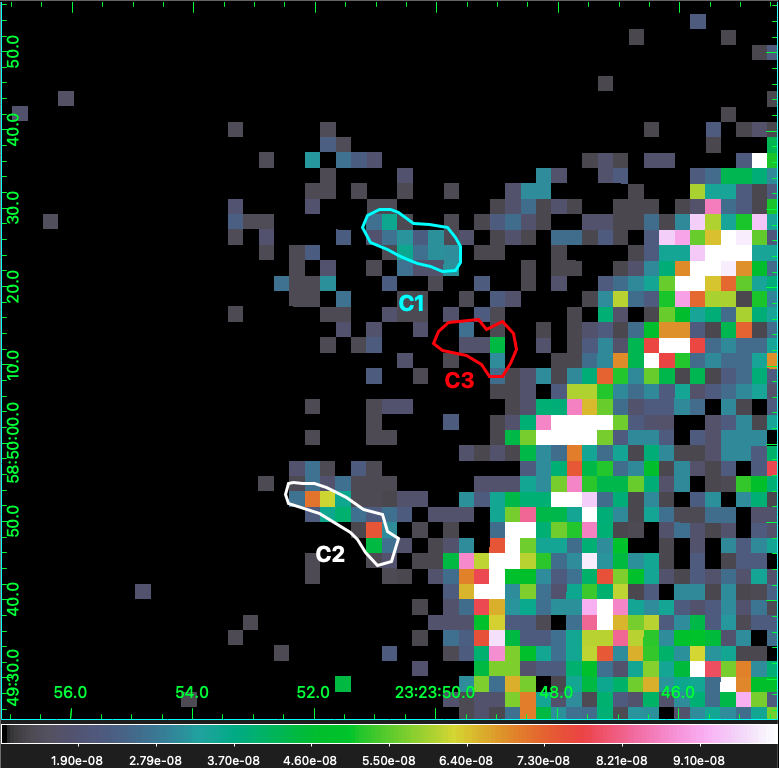}
    \includegraphics[width=\columnwidth]{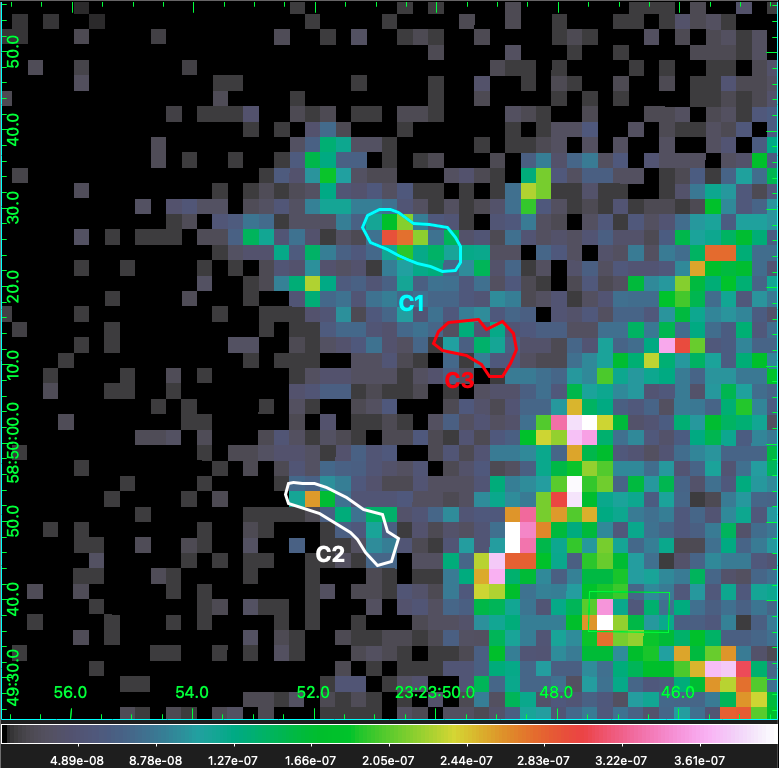}
    \caption{
    \textit{Upper left panel}: Counts image of Cas A with colorbar in logarithmic scale. The yellow box highlight the zoom region for the other panel and the green circle marks the background region used for the spectral analysis.
    \textit{Upper right panel}: Close up view of the \textit{Chandra}/ACIS flux map of Cas A in the region of the north-eastern jets in $0.7-8.0$ keV band and with the dimension of a pixel of $\sim0.5$''. 
    Source regions of the spectral analysis (S1, S2 and S3) of the different jets are marked with colored polygons.
    The color scale is linear in units of photons cm$^{-2}$ s$^{-1}$. 
    \textit{Lower panels}: Same as the \textit{upper right panel} but in the $5.5-5.8$ keV band (\textit{left panel}) and in the $6.3-6.9$ keV band (\textit{right panel}).
    The pixel size is $\sim2$'' and source regions for the detection of Cr (C1, C2 and C3) are marked with colored polygons.
    }
    \label{fig:extr}
\end{figure*}
In order to test the presence of Cr in the NE jet of Cas A, which represents a valuable tracer of the explosion conditions in CCSNe especially when co-located with other heavy elements (e.g. Iron; Fe), we created flux images in different energy bands.
In Fig. \ref{fig:extr} (\textit{lower left panel}) we show the flux map of Cas A zoomed in the NE jets.
We selected the energy band of the K-shell emission of Cr ($5.5-5.8$ keV).
For each one of the three fingers of the jet (three narrow, elongated ejecta structures, present in the NE jet, likely tracing distinct outflows produced during the explosion), we selected regions where pixels exceed by more than 20 time the value of the background ($\sim1.4\times10^{-9}$ photons cm$^{-2}$ s$^{-1}$) in the $5.5-5.8$ keV band, namely region C1, C2 and C3, marked with colored polygons in Fig. \ref{fig:extr} (\textit{lower panels}).
These regions are co-located with a flux enhancement in the Fe K-shell emission energy band (\textit{lower right panel}; $6.3-6.9$ keV). 
We extracted the spectra from these regions, and performed a spectral analysis to confirm the detection of the Cr line and verify whether the Cr abundance is enhanced with respect to the solar value.

\begin{figure*}[t!]
    \centering
    \includegraphics[width=0.9\columnwidth]{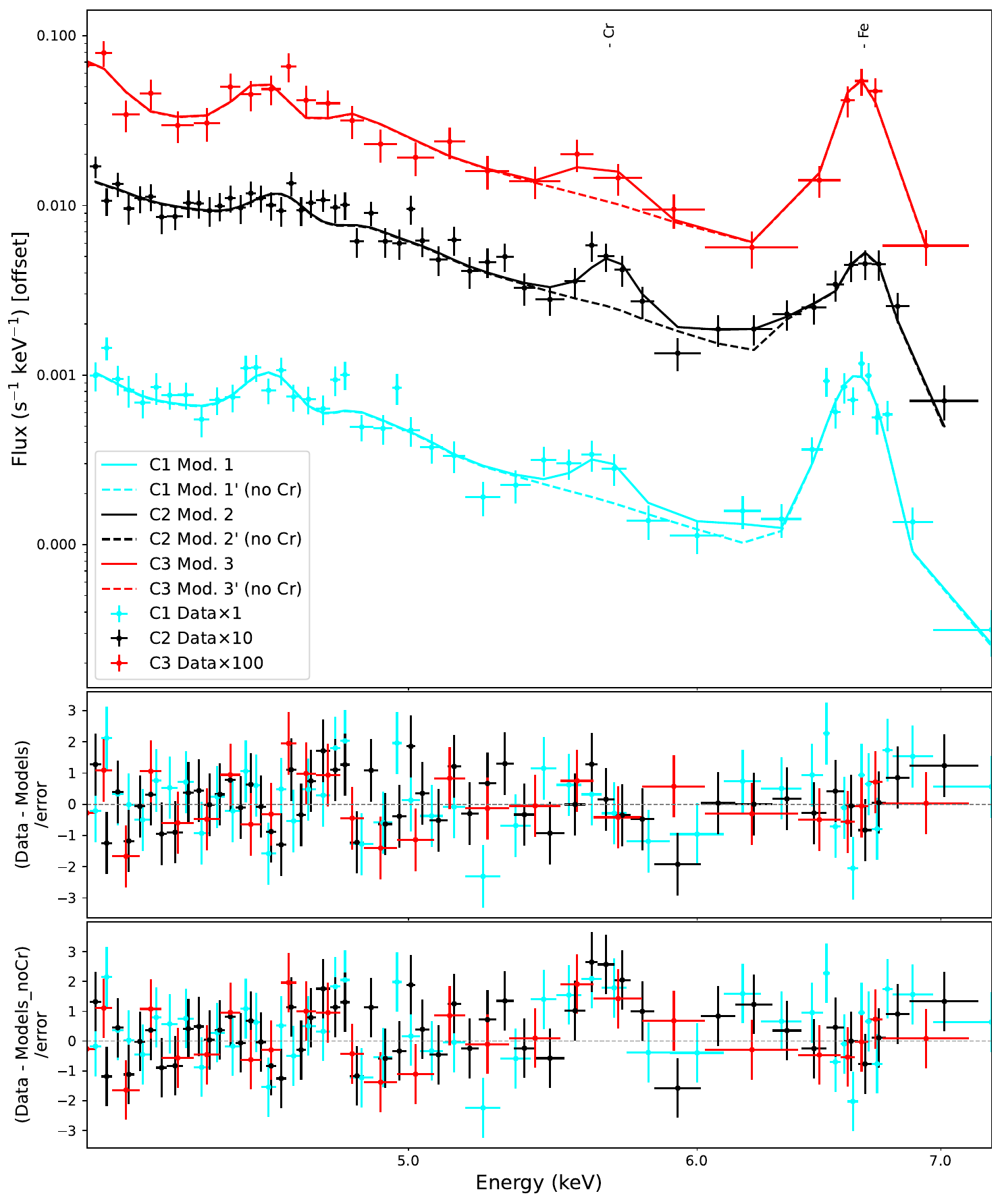}
    \includegraphics[width=1.1\columnwidth]{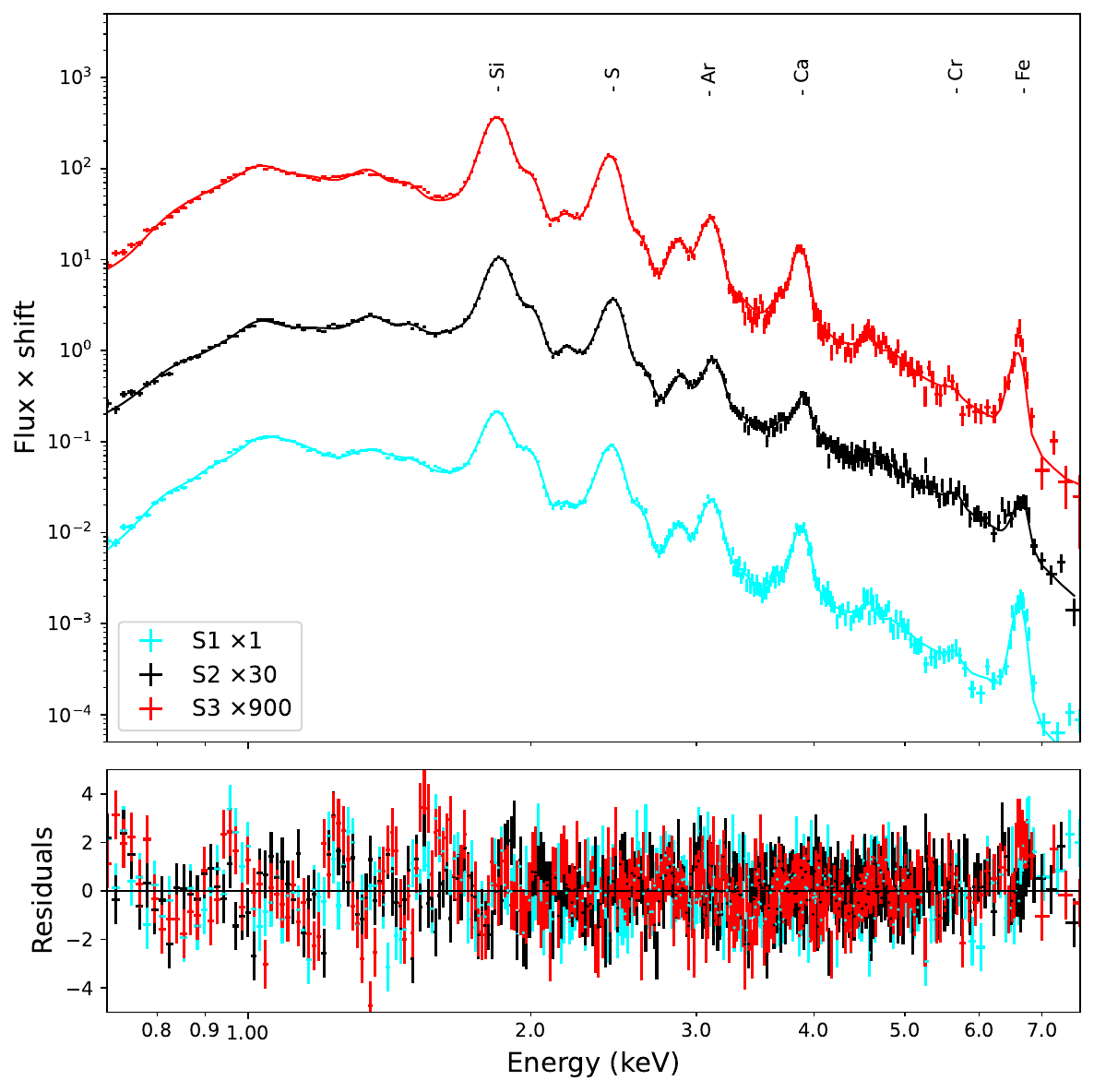}
    \caption{\textit{Left panel:} Chandra/ACIS spectra extracted from the regions shown in Fig. \ref{fig:extr} (\textit{lower panels}), with their corresponding best-fit model and residuals, obtained by letting the Cr abundance free to vary (solid lines for the models and central subplot for the residuals), and fixed to the solar value (dashed lines for the models and lower subplot for the residuals). The three spectra were multiplied by different scaling factors, as indicated in the figure label, to improve visibility.
    \textit{Right panel:} Chandra/ACIS spectra extracted from the regions shown in Fig. \ref{fig:extr} (\textit{upper panels}), with their corresponding best-fit model and residuals. The three spectra were multiplied by different scaling factors, as indicated in the figure label, to improve visibility.}
    \label{fig:Crspec}
\end{figure*}

Using the same approach as \cite{2021Natur.592..537S} we performed the analysis in the $4.1-7.1$ keV band using a single-component plane-parallel shock thermal model in non‐equilibrium ionization (NEI) absorbed by the interstellar medium (\texttt{tbabs * vvpshock}).
The free parameters in the fit are the temperature ($kT$) the abundances of Ca, Cr and Fe, the ionization parameter ($\tau$) and the normalization. Left panel of Fig. \ref{fig:Crspec} shows the spectra of regions C1, C2, C3 with their corresponding best-fit model and residuals obtained by letting the Cr abundance free to vary (solid curves and upper residuals subplot) and fixed to the solar value (dashed curves and lower residuals subplot) while best-fit values are reported in Tab. \ref{tab:bf}.
Left panel of Fig. \ref{fig:Crspec} shows a remarkable Cr line in all the spectra, especially in region C2, where the peak of the Cr line emission is almost as intense as that of the Fe K line emission. This is quite impressive, and definitely unusual, since the Cr emission is typically much dimmer, often by one to two orders of magnitude in flux, than the Fe emission both in Cas A (e. g., \citealt{2021Natur.592..537S}) and in other SNRs (e. g., \citealt{2006A&A...453..567M, 2013ApJ...766...44Y}). 
We consider this as an indication of a very large Cr/Fe abundance ratio (see Sect. \ref{sec:comp}).
We found that the quality of the fits improves when allowing the Cr abundance to vary freely, resulting in Cr abundances larger than one with significance levels of 2.7$\sigma$, 3.7$\sigma$, and 2.2$\sigma$ for regions C1, C2, and C3, respectively (derived from $\sqrt{\Delta\chi^2}$). 
The improvement in the fits was assessed using an F-test, which yielded probabilities of 0.0266, 0.0002, and 0.0331 for C1, C2, and C3, respectively, indicating that the inclusion of Cr as a free parameter provides a statistically significant improvement in all regions.

We also investigated the impact of pile-up events on the spectra, as this effect can be significant in some of the brightest regions of Cas A.
In particular, strong emission lines such as Si XIII ($\sim1.85$ keV) and S XV ($\sim2.45$ keV) could create false line features at specific energies (e.g., 2$\times 1.85$ keV $\approx3.7$ keV, 1.85 keV + 2.45 keV $\approx4.3$ keV and 3$\times$1.85 keV $\approx 5.6$ keV). 
The combination of three photons originating from the Si XIII emission line might indeed create a fake Cr K-shell line.
In the pile-up map we measured a count-rate per frame in our extraction regions is always below 0.02 (counts/frames; on average) which represents the 1\% pile-up fraction threshold. 
As a consequence we can exclude that our spectra are affected by pile-up effects.

\subsection{Probing the Ejecta Composition in the NE Jet Fingers}
\label{sec:comp}
To improve the statistics, we perform our spatially resolved spectral analysis by extracting spectra from larger regions, namely by considering  each of the three fingers of the NE jet (hereafter Region S1, S2 and S3, see Fig. \ref{fig:extr}).
All spectra were fitted in the $0.7-8.0$ keV band, with the following model in XSPEC: \texttt{tbabs*(vvpshock$_1$ + vvpshock$_2$)}.
We used two NEI components to account for possible stratification in temperature ($kT$) and ionization parameter ($\tau$).
The lower bound at 0.7 keV allows access to emission lines such as the Fe L complex, which empower us to better constrain $kT$, $\tau$ and the Fe abundance, while the upper bound at 8.0 keV corresponds to the effective high-energy limit of the \textit{Chandra} ACIS instrument.
The abundances of Mg, Si, S, Ar, Ca, Cr and Fe were left free to vary, and linked to be equal in the two thermal components in NEI.
A power law component was added when fitting the spectrum of region S2 to account for a residual non-thermal synchrotron emission.
The best-fit parameters are listed in Tab. \ref{tab:bf}, while the spectra with the best-fit model and residuals are shown in Fig. \ref{fig:Crspec} (\textit{right panel}).

We first notice that the chemical composition retrieved for region S1, S2, S3 is consistent with that of the corresponding subregion C1, C2, C3, respectively, but with much smaller error bars. This shows that the chemical composition is consistent with being uniform within each of the three fingers, while variations can be noticed when comparing different fingers.
\begin{table*}[htbp]
\centering
\caption{Best-fit parameters fo region C1, C2, C3, S1, S2 e S3 shown in Fig. \ref{fig:extr}.}
\label{tab:bf}
\begin{tabular}{lcccccc}
\hline\hline
\textbf{Parameter} & \textbf{C1} & \textbf{C2} & \textbf{C3} & \textbf{S1} & \textbf{S2} & \textbf{S3} \\
\hline
$n_{\rm H}\ (10^{22}\,\mathrm{cm^{-2}})$ &
  \multicolumn{3}{c}{1.4 (fixed)}&
  $1.64_{-0.03}^{+0.06}$ &
  $1.41_{-0.04}^{+0.02}$ &
  $1.389_{-0.017}^{+0.026}$ \\
\hline
\multicolumn{7}{c}{\textbf{vvpshock$_1$}} \\
\hline

$kT\ (\mathrm{keV})$ &—&—&—&
  $0.277_{-0.015}^{+0.031}$ &
  $1.14_{-0.20}^{+0.25}$ &
  $0.755_{-0.016}^{+0.023}$ \\

$\tau_u\ (10^{11}\,\mathrm{s/cm^3})$ &—&—&—&
  10 (fixed) & 
  $2.0_{-0.8}^{+1.5}$ &
  $9.8_{-2.0}^{+8.2}$ \\

norm$_{kT1}$  $(10^{-5})$ &—&—&—&
  $37_{-10}^{+9}$ &
  $4.2_{-1.1}^{+1.2}$ &
  $4.6_{-0.5}^{+0.4}$ \\
\hline
\multicolumn{7}{c}{\textbf{vvpshock$_2$}} \\
\hline

$kT\ (\mathrm{keV})$ &
  $2.5_{-0.8}^{+3.}$&
  $2.1_{-0.6}^{+0.9}$&
  $2.6_{-1.0}^{+3.3}$&
  $2.33_{-0.06}^{+0.22}$ &
  $3.6_{-0.5}^{+0.7}$ &
  $3.3_{-0.2}^{+0.3}$ \\

[Mg]/[Mg$_{\astrosun}$] &—&—&—&
  $3.6_{-1.0}^{+0.8}$ &
  $5.13\pm0.36$ &
  $14.9_{-0.6}^{+1.2}$ \\

[Si]/[Si$_{\astrosun}$] &—&—&—&
  $43_{-2}^{+3}$ &
  $52_{-7}^{+10}$ &
  $95_{-4}^{+7}$ \\

[S]/[S$_{\astrosun}$] &—&—&—&
  $54\pm3$ &
  $58_{-11}^{+7}$ &
  $133_{-6}^{+13}$ \\

[Ar]/[Ar$_{\astrosun}$] &—&—&—&
  $57_{-4}^{+5}$ &
  $54_{-8}^{+3}$ &
  $133_{-6}^{+13}$ \\

[Ca]/[Ca$_{\astrosun}$] &
  $140_{-60}^{+110}$&
  $120_{-60}^{+90}$&
  $280_{-170}^{+620}$ &
  $89_{-8}^{+10}$ &
  $62_{-6}^{+5}$ &
  $220_{-13}^{+31}$ \\

[Cr]/[Cr$_{\astrosun}$] &
  $100_{-80}^{+180}$&
  $150_{-90}^{+150}$&
  $220_{-190}^{+580}$&
  $60_{-50}^{+50}$ &
  $90_{-50}^{+50}$ &
  $<150$ \\

[Fe]/[Fe$_{\astrosun}$] &
  $25_{-15}^{31}$&
  $10_{-4}^{+7}$&
  $38_{-20}^{+70}$&
  $25_{-5}^{+3}$ &
  $8.8_{-1.5}^{+1.2}$ &
  $27.8_{-1.9}^{+3.9}$ \\

$\tau_u\ (10^{11}\,\mathrm{s/cm^3})$ &
  $2.6_{-1.2}^{+3.8}$&
  $7\pm4$&
  $5_{-2}^{+5}$&
  $1.94_{-0.15}^{+0.14}$ &
  $1.13_{-0.09}^{+0.17}$ &
  $1.06_{-0.08}^{+0.06}$ \\

Redshift ($10^{-4}$) &
  —&
  —&
  —&
  $-4_{-2}^{+10}$ &
  $-68.78_{-0.02}^{+4.23}$ &
  $6.12_{-0.07}^{+0.74}$ \\

norm$_{kT2}$ ($10^{-5}$) &
  $2.5_{-3}^{+2}$&
  $3.1_{-0.7}^{1.0}$&
  $0.58_{-0.16}^{+0.23}$&
  $6.6_{-0.4}^{+0.3}$ &
  $4.0_{-1.4}^{+1.0}$ &
  $2.5\pm 0.2$ \\
\hline
\multicolumn{7}{c}{\textbf{powerlaw}} \\
\hline
Photon index $\Gamma$ &—&—&—&
  — &
  $3.31_{-0.18}^{+0.12}$ &
  — \\
norm$_{\rm pl}\ (10^{-4} $ keV$^{-1}$ cm$^{-2}$ s$^{-1}$) &—&—&—&
  — &
  $3.32_{-0.08}^{+0.06}$ &
  — \\
\hline
\hline
\multicolumn{7}{c}{\textbf{Mass Ratios}} \\
\hline
Si/Fe &—&—&—& $0.63_{-0.12}^{+0.09}$ & $2.2\pm0.5$ & $1.303_{-0.104}^{+0.209}$ \\
S/Fe  &—&—&—& $0.41_{-0.08}^{+0.06}$ & $1.3_{-0.3}^{+0.2}$ & $0.953_{-0.084}^{+0.148}$ \\
Ar/Fe &—&—&—& $0.123_{-0.025}^{+0.018}$ & $0.34_{-0.08}^{+0.05}$ & $0.266_{-0.021}^{+0.046}$ \\
Ca/Fe &$0.20_{-0.14}^{+0.29}$&$0.42_{-0.26}^{+0.45}$&$0.25_{-0.21}^{+0.74}$& $0.12_{-0.03}^{+0.02}$ & $0.25_{-0.05}^{+0.04}$ & $0.277_{-0.025}^{+0.055}$ \\
Cr/Fe &$<0.08$&$0.14_{-0.10}^{+0.18}$&$<0.17$& $0.021\pm0.017$ & $0.09\pm0.06$ & $<0.05$ \\
\hline
\hline 
\(\chi^2/\mathrm{d.o.f.}\) &
  $56.78/39$ &
  $45.35/49$ &
  $19.14/19$&
  $418.65/309$ &
  $381.12/339$ &
  $538.37/306$ \\
\hline
\end{tabular}
\begin{threeparttable}
\begin{tablenotes}
  \item \textbf{Notes.} Errors are at the $2\sigma$ confidence level and approximated to the first significant digit. Normalization parameters for the \texttt{vvpshock} components are defined as $\frac{10^{-14}}{4 \pi D^2}\int n_\mathrm{e} n_\mathrm{H} dV$ (cm$^{-5}$) where $D$ is the angular distance to the source, $dV$ is the volume element and $n_{\rm e}$ and $n_{\rm H}$ are the electron and Hydrogen densities, respectively.
  \end{tablenotes}
  \end{threeparttable}
\end{table*}

The best-fit models successfully reproduce the overall continuum and line emission features observed in the $0.7–8.0$ keV range. 
In particular, strong emission lines from Si, S, Ar, Ca, and Fe are well matched by the model as well as the Cr emission line.
Table \ref{tab:bf} shows that, while in region S1 and S3 we measure very low values of the Doppler shift (consistent with these to fingers to lie almost in the plane of the sky), the spectrum of region S2 shows a strong blueshift, which corresponds  to a velocity of about $2100$ km s$^{-1}$, directed towards the observer.

The need to adopt two components to model the spectra is indicative of a distribution of temperatures and ionization parameters in the ejecta: \texttt{vvpshock}$_1$ component reflects a cooler plasma, closer to the equilibrium of ionization, while \texttt{vvpshock}$_2$ shows higher temperatures, and lower values of the ionization parameters.
Residuals are generally low and randomly distributed, suggesting no significant systematic deviations. 
Minor discrepancies remain around the Fe-L and Mg region, but they do not significantly impact the abundance estimates.
In the Fe-K region of the S3 spectrum, the model slightly underestimate the line flux, and there may be a small energy offset.
Despite this, the overall fit remains statistically acceptable and the derived Fe abundance is robust, as the fit is primarily driven by the Fe-L lines complex which for this region is well modeled.


Table \ref{tab:bf} shows that large values of the Cr/Fe abundance ratios are retrieved, especially in region S2, where we measure the largest Cr/Fe abundance ratio with the lowest Fe abundance. 
Since CCD spectral fits exhibit a strong degeneracy between emission measure and absolute abundances, allowing us to reliably derive only mass (or abundance) ratios rather than absolute mass values \citep{2020A&A...638A.101G}, for each region, we computed mass ratios of selected species (namely, Si, S, Ar, Ca, Cr) with respect to Fe as:
\begin{equation}
\frac{M_\mathrm{X}}{M_{\mathrm{Fe}}}= \frac{A_\mathrm{X} A_{\mathrm{X}_{\odot}}\mu_\mathrm{X}}{A_\mathrm{Fe} A_{\mathrm{Fe}_{\odot}}\mu_{\mathrm{Fe}}},
\end{equation}
where $A_\mathrm{X}$ is the best‐fit abundance of element X relative to the solar value (i.e. the parameter returned by the \texttt{vvpshock} models in the fit), $A_{\mathrm{X}_{\odot}}$ is the corresponding solar abundance (taken from \citealt{1989GeCoA..53..197A}), and $\mu_\mathrm{X}$ is the atomic mass number of element X.
The computed mass ratios for each region are reported in Tab. \ref{tab:bf}.

\section{Discussions and Conclusion}\label{sect:con}
\subsection{Cr-K Detection and Morphology}
Our deep Chandra/ACIS imaging analysis of the NE jet reveals a clear structure in the band of the K‑shell line from helium‑like Cr co‑spatial with the Fe‑K emission at the ``basis" of the fingers (Fig. \ref{fig:extr}, lower panels).
Our spectral analysis confirms the presence of a particularly strong Cr K$\alpha$ emission.

Based on the best-fit abundances, we estimate Cr/Fe mass ratios of $\sim$0.02 in S1, $\sim$0.1 in S2, and $<$0.05 in S3. 
Among them, Region S2 clearly stands out, with a Cr/Fe value approximately four times higher than in the adjacent fingers. 
This region indeed exhibits an high Cr abundance, while showing the lowest Fe abundance among the three fingers, resulting in the largest Cr/Fe mass ratio ever reported in Cas A.

By comparison, \cite{2021Natur.592..537S} detected Cr K$\alpha$ in the southeastern Fe‑rich plumes of Cas A but measured much lower Cr/Fe mass ratios (only 0.004–0.007 at 99\% confidence level) reflecting the dominance of iron in those regions. 
Likewise, \cite{2022PASJ...74..334I} extracted spectra from all three northeast jet fingers as a single, combined region. 
This approach effectively dilutes the localized Cr emission we identify in S2, producing a substantially lower overall Cr/Fe ratio than the peak value measured here ($\mathrm{Cr_{S2}/Fe_{S2}} \sim 0.1$ vs. $\mathrm{Cr_{Ikeda22}/Fe_{Ikeda22}} \sim 0.04$). 
In contrast, our targeted analysis of the three different fingers isolates the zone of maximal Cr emission in region S2, revealing the highest Cr/Fe enhancement yet seen in Cas A.

\subsection{Constraining the Origin of the Three Fingers}
\subsubsection{Nucleosynthetic Model description}
We compare the observed elemental abundance patterns in the NE jet regions with theoretical yields of CCSN nucleosynthesis models utilized by \cite{2020ApJ...893...49S,2021Natur.592..537S} to address the origin of the three fingers in the NE jet of Cas A.
The evolution of the $15\,M_{\odot}$ progenitor star with a sub-solar metallicity ($0.5 \,Z_{\odot}$) is calculated from the hydrogen burning phase until just before the core-collapse (the central temperature reaches 10$^{9.9}$ K) by a one-dimensional stellar evolution code, HOngo Stellar Hydrodynamics Investigator \citep[HOSHI:][]{2016MNRAS.456.1320T,2018ApJ...857..111T,2019ApJ...871..153T,2019ApJ...881...16Y}. 
CCSN explosions are carried out by injecting thermal energy (the so-called thermal bomb) in the vicinity of the iron-silicon interface of the progenitor star, with a one-dimensional hydrodynamic code \citep{2005ApJ...619..427U}. The injection energies are adjusted to obtain an explosion energy $E_{\rm exp}$. 
The four cases of $E_{\rm exp}=$ (1, 2, 3, 5) $\times$10$^{51}$ erg are calculated, speculating that the corresponding (effective) explosion energy can vary depending on the region. 
The location of the mass cut, i.e. the mass coordinate that separates the fall-back material from the ejecta, is determined as the ejected $^{56}$Ni mass to be $0.07\,M_{\odot}$. 
Nucleosynthesis calculations are performed based on the one-dimensional hydrodynamic models; radioactive isotopes are decayed until the age of Cas A ($\sim 350$ yr) to be compared with the observation.
Since the different explosion energies result in qualitatively similar features, we present only the case of $E_{\rm exp} = 3\times10^{51}$ erg, as in \cite{2021Natur.592..537S}.
This choice is further supported by estimates of the explosion energy of Cas A \citep{2003ApJ...597..347L, 2003ApJ...597..362H}, which range between $2$ and $3\times10^{51}$ erg.

\subsubsection{Comparison between model and mass ratios}
\begin{figure*}[t!]
    \centering
    \includegraphics[width=1.\linewidth]{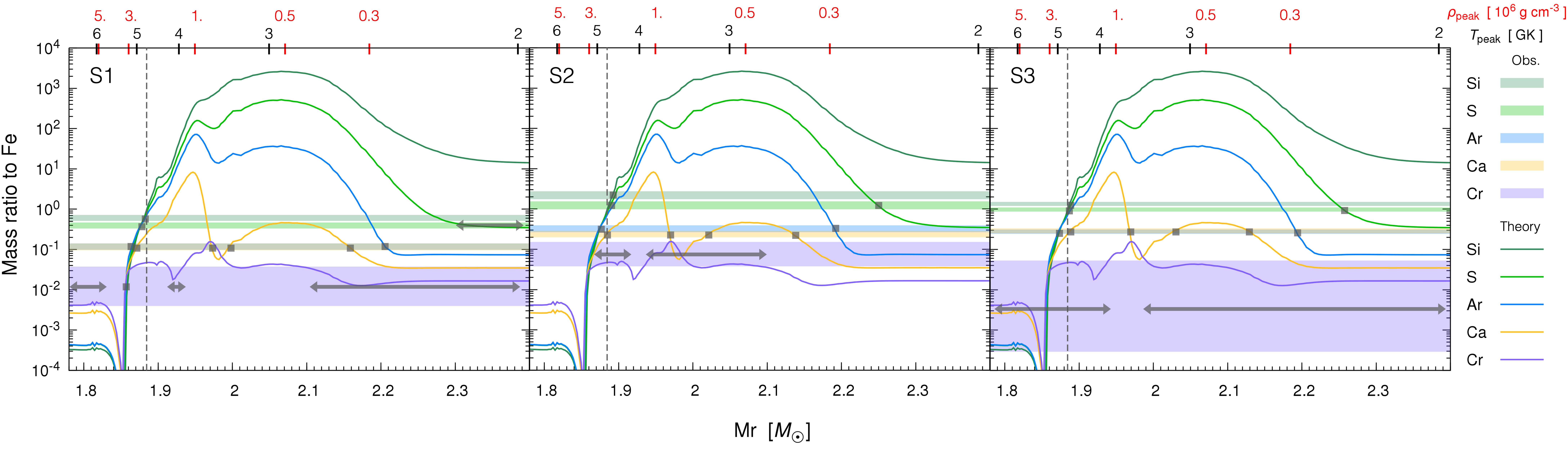}
    \caption{Mass fractions of Si, S, Ar, Ca, and Cr with respect to Fe in the one-dimensional CCSN nucleosynthesis model with $E_{\rm exp}=3\times10^{51}$ erg as a function of the Lagrangian mass coordinates (Mr). The shaded areas are the range ($2\sigma$ confidence level) of mass fractions measured by our spectral analysis in region S1, S2 and S3 (\textit{left, center}, and \textit{right} panels, respectively). The colors of shaded regions are semi-transparent; overlapped areas are shown as the color composites. Gray arrows and square points mark where the model and the observation match. 
    The red and black ticks on the upper border indicate the corresponding peak densities and temperatures, respectively.
    The vertical dashed lines denote the transition points between the iron- and silicon-dominated regions, which approximately separate the complete Si-burning via $\alpha$-rich freeze-out from the incomplete Si-burning with quasi-statistical equilibrium (QSE).
    In regions S1 and S3, the observed ranges of Ar and Ca are almost overlapped with each other.}
    \label{fig:compS2}
\end{figure*}
Figure \ref{fig:compS2} shows a direct comparison between the measured mass ratios and all the configurations calculated as a function of the Lagrangian mass coordinates (Mr). The full set of mass ratios observed in regions S1, S2, and S3, cannot be exactly reproduced by any of the model mesh points (within the 2$\sigma$ confidence level). 
Despite the presence of some isolated model–data matches outside the transition region between complete and incomplete Si burning (marked by the vertical dashed line in Fig. \ref{fig:compS2}) we notice that, remarkably, the three regions exhibit abundance patterns that are nearly consistent with each others in the proximity of the transition zone.
When considering only Si and S one finds a reasonable match with mesh points in the incomplete Si-burning regime for region S2.
Such conditions correspond to a narrow shell in the progenitor where the peak temperature ($>4.5$ GK) is high enough to activate quasi-statistical equilibrium (QSE) reactions but not sufficient for full $\alpha$-rich freeze-out.
Conversely, when focusing on Ar and Ca, the best match shifts to deeper, hotter regions associated with complete Si burning. 
Region S3 occupies an intermediate position, with elemental mass coordinates that straddle the boundary between incomplete and complete Si-burning. 
In contrast, region S1 lies fully within the complete Si-burning zone, where the higher peak temperatures result in more advanced nuclear processing, including significant $\alpha$-rich freeze-out.

In general, the discrepancy between the different mass ratios suggests that the ejecta of the NE jet did not originate from a single homogeneous burning layer but rather experienced post-explosive hydrodynamic mixing. 
Adjacent burning zones mixing, possibly driven by Rayleigh-Taylor instabilities during the early phases of the explosion, could reconcile the observed composition.
To explore whether the observed abundance patterns in the NE jet fingers could be explained by a combination of different nucleosynthetic layers, we tested a physically motivated two-zone mixing model based on the Lagrangian mass coordinates of the progenitor, as a simple example.
We adopted the two-zone mixing approach because it allows for the mixing of non-adjacent layers, as expected in realistic three-dimensional instabilities, while remaining conceptually straightforward and computationally convenient.

\begin{figure}[t!]
    \centering
    \includegraphics[trim={0 0 10pt 10pt},clip,width=\linewidth]{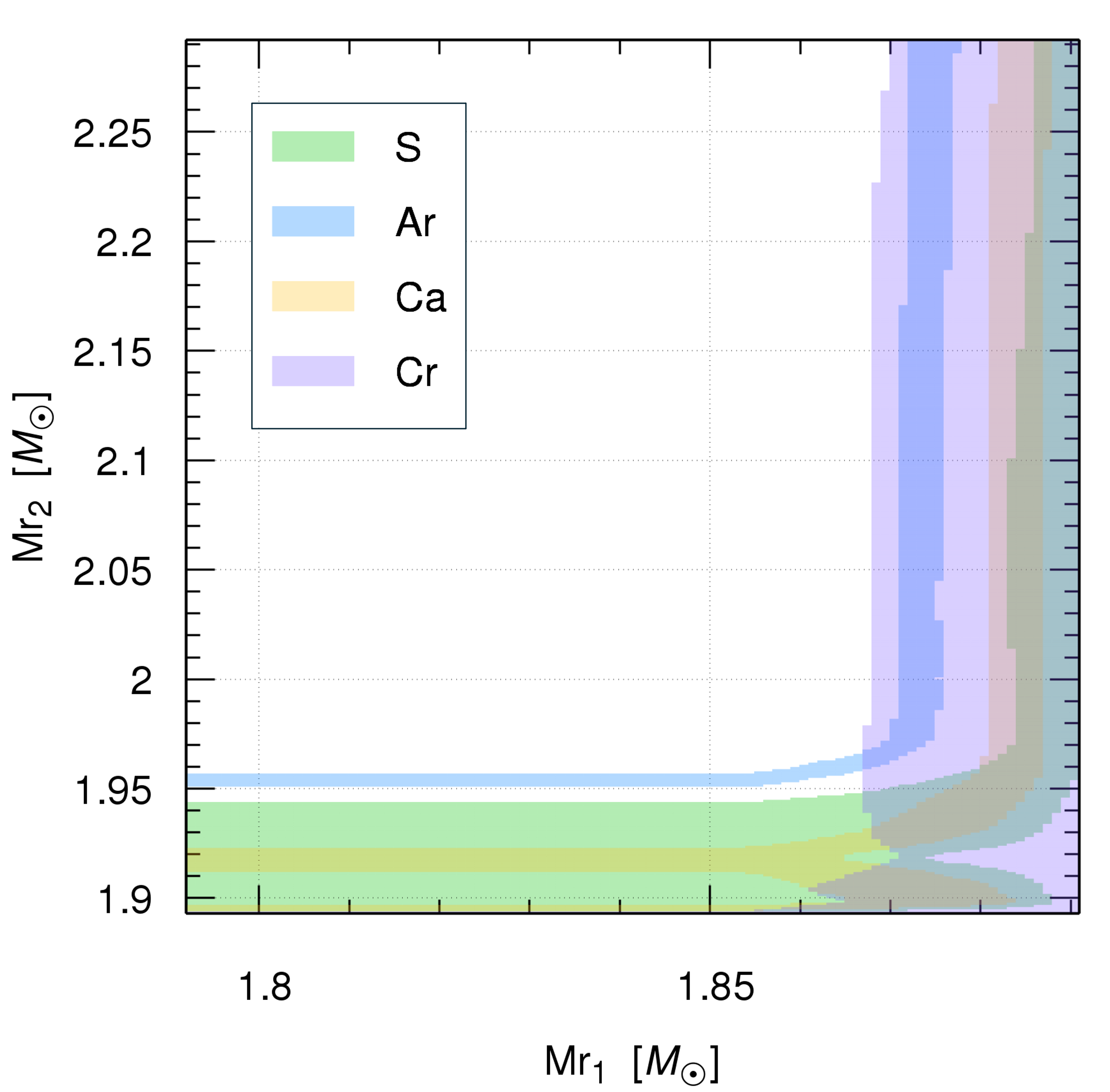}
    \caption{
    The combinations of two Lagrangian points consistent with the observed mass ratios for the region S2 are shown as shaded colors for S, Ar, Ca, and Cr. The relevant mass fractions are derived by the average of two Lagrangian points with weights recovering the observed Si/Fe ratio. The horizontal (Mr$_1$) and vertical (Mr$_2$) axes denote the mass coordinates of the first (inner) and the second (outer) Lagrangian points, respectively. Then, the regions in the two-dimensional grid consistent with the observed 2$\sigma$ ranges are shown. The colors of shaded regions are semi-transparent; overlapped areas are shown as the color composites.}
    \label{fig:two-zone}
\end{figure}

We first identified the Lagrangian coordinate that reproduces the best-fit Si/Fe ratio for each region.
Then, we selected pairs of Lagrangian points on either side of this boundary—one from a deeper (with mass coordinates Mr$_1$), hotter layer and one from an outer (with mass coordinates Mr$_2$), cooler region.
Their corresponding mass fractions were averaged with weights chosen to recover the observed Si/Fe ratio. 
The combinations of two Lagrangian points consistent with the observed mass ratios for the case of region S2 is shown in Figure~\ref{fig:two-zone} as a representative example.

In this scenario, S1, S2, and S3 would sample different degrees of mixing across the boundary between the complete and incomplete Si-burning regions.
The evidence of mixing has critical implications for both the explosion dynamics and the geometry of the ejecta. 
It supports a model in which the NE jet is not a clean, layered outflow but a chemically inhomogeneous structure shaped by multidimensional instabilities during shock propagation through the composition interfaces of the progenitor star and/or shock breakout and reverse-shock interaction. 
This complexity is essential to reproduce the diversity of abundance patterns observed across the three jet regions.
This created a two-dimensional parameter space of possible mixing combinations, where we evaluated whether the resulting mass ratios matched our observations.
However no single pair of mixing points was able to reproduce all observed mass ratios simultaneously. 
In particular, the mass ratios of S, Ca, and Cr to Fe can be simultaneously consistent with the observations, however, the Ar/Fe ratio remained consistently incompatible with the parameter space that matched the other elements except for the Cr/Fe. 
This discrepancy persisted even when using equal weights, suggesting that Ar is synthesized under thermodynamic conditions not easily reconcilable with the rest of the observed composition through simple two-zone mixing. 

The relatively stable behavior of Ar (mass ratios of Ar to others) across burning layers in the model may be responsible for this tension.
These results indicate that while a degree of mixing is clearly required to reproduce the global abundance patterns observed in the NE jet, a more complex scenario involving multi-zone mixing or non-standard thermodynamic trajectories (e.g., enhanced electron fraction (Ye), extreme entropy conditions) may be necessary to fully account for the observed elemental ratios. 
Our analysis confirms that the jet ejecta do not originate from a single, well-defined burning layer, but rather reflect a chemically stratified structure shaped by hydrodynamic instabilities during the explosion.

\subsubsection{Dynamics and Differencies of the Three Fingers}
While the three fingers in the NE jet share broad similarities in morphology and overall thermal structure, they show marked differences in both their kinematics and chemical composition, indicating a complex and asymmetric origin.
Indeed region S2 shows an exceptionally high Cr/Fe mass ratio ($\sim0.1$) among the three fingers.
Also, comparing the chemical composition of this region with the nucleosynthesis models it appears that regions S2 originated from a slightly external ejecta layer, close to the boundary between complete and incomplete Si-burning. 
Conversely, S1 and S3 show higher Fe abundances, lower Cr/Fe ratios and the comparison with the model indicates a deeper origin within the star, potentially incorporating a greater fraction of complete Si-burning material.

Remarkably, these chemical differences are accompanied by noticeable variations in the dynamics of the 3 fingers. 
We indeed derived the radial velocities of the plasma in the three fingers from our spectral fitting.
Regions S1 and S3 show a modest to negligible Doppler shift (see Tab. \ref{tab:bf}) leading to a line of sight (LoS) velocity of $\sim 60$ km s$^{-1}$ toward the observer for Region S1 and $\sim 180$ km s$^{-1}$ away from the observer for region S3. On the other hand, region S2 shows a substantial blue-shift, with a LoS velocity of $\sim 2100$ km s$^{-1}$ toward the observer).
Using the value of 7200 km s$^{-1}$ for the velocity on the plane of the sky from \cite{2024ApJ...974..245S}, this suggest that S2 is tilted toward the observer of $\sim16$\textdegree, in contrast to S3 and S1 which lie closer to the plane of the sky.

The combined kinematic and compositional signatures indicate that the three fingers, while part of a coherent NE outflow, are not homogeneous neither co-spatial. 
Their differences point to a jet-like structure shaped by both explosion asymmetry and mixing processes, where material from different depths and burning conditions was channeled into slightly divergent trajectories.

\subsection{Summary and Conclusions}
In this work, we presented a spatially resolved X-ray spectral analysis of the NE jet of Cas A using the deepest available Chandra/ACIS dataset. 
Focusing on the three individual fingers, we characterized their thermal, chemical, and dynamical properties in detail, revealing important clues about their origin and the explosion mechanism.

We report the detection of Cr-K$\alpha$ emission, with the strongest signal emerging from the southernmost finger (C2 and S2 regions). 
This finger of the NE jet shows the highest Cr/Fe mass ratio ever measured in Cas A ($\sim0.14$), in stark contrast to previous studies focused on Fe-dominated ejecta or on broader areas. 
This extreme mass ratio, together with the chemical composition of this finger, points toward the boundary between the complete and the incomplete Si-burning as the region where this structure originated. 
Indeed, comparisons with 1D nucleosynthesis models show that no single burning layer can simultaneously reproduce all the measured abundance ratios across the three regions. 
By considering the model-observation matches which cluster near the transition layer, Si and S are broadly consistent with incomplete Si-burning, while Ar and Ca favor deeper, hotter conditions. 

This mismatch strongly suggests post-explosive mixing, likely driven by Rayleigh-Taylor instabilities, as a way to reconcile the observed patterns. 
All three fingers seem to sample the boundary zone between complete and incomplete Si-burning, but with different degrees of chemical mixing and different depth in burning regimes. In particular, the central and northernmost fingers of the NE jet likely originated in deeper layers than region S2. 

We also reconstructed the 3D velocity vectors of the three fingers using LoS velocities from redshift measurements and velocities on the plane of the sky from \cite{2024ApJ...974..245S}. 
Region S2 stands out not only for its chemical uniqueness, but also for its kinematics: it is significantly blueshifted ($\sim 2100$ km s$^{-1}$) and tilted toward the observer of about 16\textdegree, while S1 and S3 lie closer to the plane of the sky. 
These distinctions in both composition and motion implies that the jet is not a uniform outflow, but rather a chemically and dynamically stratified structure, shaped by directional asymmetries and multi-dimensional hydrodynamics during the explosion.

Altogether, our findings support a scenario in which the NE jet in Cas A channels material from a range of nucleosynthetic layers, with evidence for strong internal mixing and directional asymmetry. 
The detection of Cr-rich, Fe-poor knots offers critical constraints for explosion models aiming to reproduce the complex geometry and composition of Cas A.

\begin{acknowledgments}
V.S., S.O. and M.M. acknowledge financial contribution from the PRIN MUR “Life, death and after-death of massive stars: reconstructing the path from the pre-supernova evolution to the supernova remnant” funded by European Union - Next Generation EU and from the INAF Theory Grant ``Supernova remnants as probes for the structure and mass-loss history of the progenitor systems''.
E.G. and V.S. acknowledge support from the INAF Minigrant RSN4 "Investigating magnetic turbulence in young Supernova Remnants through X-ray observations".
S.N. is supported by JSPS Grant-in-Aid for Scientific Research (KAKENHI) (A), Grant Number JP25H00675, and the JST ASPIRE Program ‘RIKEN-Berkeley Mathematical Quantum Science Initiative.
\end{acknowledgments}

%
\facilities{This paper employs a list of Chandra datasets, obtained by the Chandra X-ray Observatory, contained in the Chandra Data Collection ~\dataset[DOI: 10.25574/cdc.209
]{https://doi.org/10.25574/cdc.209}}

\software{\textit{CIAO} \citep{2006SPIE.6270E..1VF}}


\bibliography{sample7}{}
\bibliographystyle{aasjournalv7}
\end{document}